\documentclass[conference]{IEEEtran}
\IEEEoverridecommandlockouts
\usepackage{cite}
\usepackage{amsmath}
\usepackage[linesnumbered,ruled]{algorithm2e}
\usepackage{graphicx}
\usepackage{amssymb}
\usepackage{textcomp}
\usepackage{xcolor}
\def\BibTeX{{\rm B\kern-.05em{\sc i\kern-.025em b}\kern-.08em
    T\kern-.1667em\lower.7ex\hbox{E}\kern-.125emX}}

\begin{document}

\title{Fast and Scalable Memristive In-Memory Sorting with Column-Skipping Algorithm\\
\thanks{*Corresponding authors}
}

\author{\IEEEauthorblockN{Lianfeng Yu, Zhaokun Jing, Yuchao Yang$^{*}$, Yaoyu Tao$^{*}$}
\IEEEauthorblockA{\textit{School of Integrated Circuits, Peking University} \\
Beijing, China 100871 \\
\{ylfwind, jingzk, yuchaoyang, taoyaoyutyy\}@pku.edu.cn}
}

\maketitle

\begin{abstract}

Memristive in-memory sorting has been proposed recently to improve hardware sorting efficiency. Using iterative in-memory min computations, data movements between memory and external processing units can be eliminated for improved latency and energy efficiency. However, the bit-traversal algorithm to search the min requires a large number of column reads on memristive memory. In this work, we propose a column-skipping algorithm with help of a near-memory circuit. Redundant column reads can be skipped based on recorded states for improved latency and hardware efficiency. To enhance the scalability, we develop a multi-bank management that enables column-skipping for dataset stored in different memristive memory banks. Prototype column-skipping sorters are implemented with a 1T1R memristive memory in 40nm CMOS technology. Experimented on a variety of sorting datasets, the length-1024 32-bit column-skipping sorter with state recording of 2 demonstrates up to 4.08$\times$ speedup, 3.14$\times$ area efficiency and 3.39$\times$ energy efficiency, respectively, over the latest memristive in-memory sorting.

\end{abstract}


\section{Introduction}

The past decade has witnessed an explosive growth of data and the needs for high-speed data processing. A large-scale data often needs to be sorted to enable higher efficiency. Sorting is a key kernel in many applications such as data mining \cite{aggarwal2015data}, robotics \cite{bayindir2016review} and machine learning \cite{devlin2018bert}. To efficiently sort an array into an order, numerous sorting algorithms have been invented in the past, such as merge-sort \cite{goldstine1947planning} or quick-sort \cite{quicksort}. These algorithms can be accelerated using CPUs/GPUs \cite{chhugani2008efficient,zhang2016high,satish2009designing}, FPGAs \cite{chen2017computer,chen2019sorting,samardzic2020bonsai,song2016parallel} and ASICs \cite{norollah2019rths,najafi2018low,lin2017hardware}. However, transferring data between memory and external processing units incurs a long latency and a degraded energy efficiency. Techniques like memory management \cite{stehle2017memory} have been developed to minimize the data movement, but such optimizations do not fundamentally solve the problem.

Memristive in-memory sorting \cite{rram_sort_alam2020,prasad2021memristive} have been proposed recently to tackle this challenge. Memristor-aided logic is developed in \cite{rram_sort_alam2020} to implement compare-and-select blocks in memory. However, a large number of memristor cells are used to implement logic gates with frequent write operations, resulting in a low memory density and a degraded device lifetime. The latest memristive in-memory sorting \cite{prasad2021memristive} uses iterative in-memory min computations with help of a near-memory circuit. The min values are searched by traversing each bit column using column reads (CR) on a 1T1R memristive memory. Frequent write operations in \cite{rram_sort_alam2020} are eliminated; however, the number of CRs is proportional to the number of 1T1R cells in the memristive memory, degrading the latency and energy efficiency.

In this work, we propose a column-skipping algorithm to minimize the number of CRs for improved sorting speed and hardware efficiency. A near-memory circuit is designed to keep track the column read conditions and skip those that are leading 0's or have been processed previously. A multi-bank management is developed to enhance the scalability when sorting a larger array stored in different memristive memory banks. Implemented in a 40nm CMOS technology with 1T1R memristive memory and experimented on a variety of sorting datasets, the length-1024 32-bit column-skipping memristive sorter with state recording of 2 demonstrates up to 4.08$\times$ speedup, 3.14$\times$ area efficiency and 3.39$\times$ energy efficiency, respectively, over the latest memristive in-memory sorting implementation \cite{prasad2021memristive}. 

\section{Background}

\subsection{Sorting Applications}

Sorting is a known bottleneck for many applications \cite{aggarwal2015data,bayindir2016review,devlin2018bert}. Here we briefly introduce two representative applications where sorting dominates the execution time: 1) Kruskal's algorithm for minimum spanning tree (MST). In Kruskal's algorithm, all the graph edges need to be sorted from low weight to high weight. Majority of the weights are small numbers with frequent repetitions; 2) MapReduce in distributed systems. In MapReduce, maps need to be sorted before transferring to the reducer stage \cite{dean2008mapreduce}. These maps are typically clustered in a few groups. We use datasets generated from these two applications for benchmarking in Section~\ref{section:evaluation}.

\subsection{Memristive In-Memory Sorting}

\begin{figure}
    \centering
    \includegraphics[width = 0.95\linewidth]{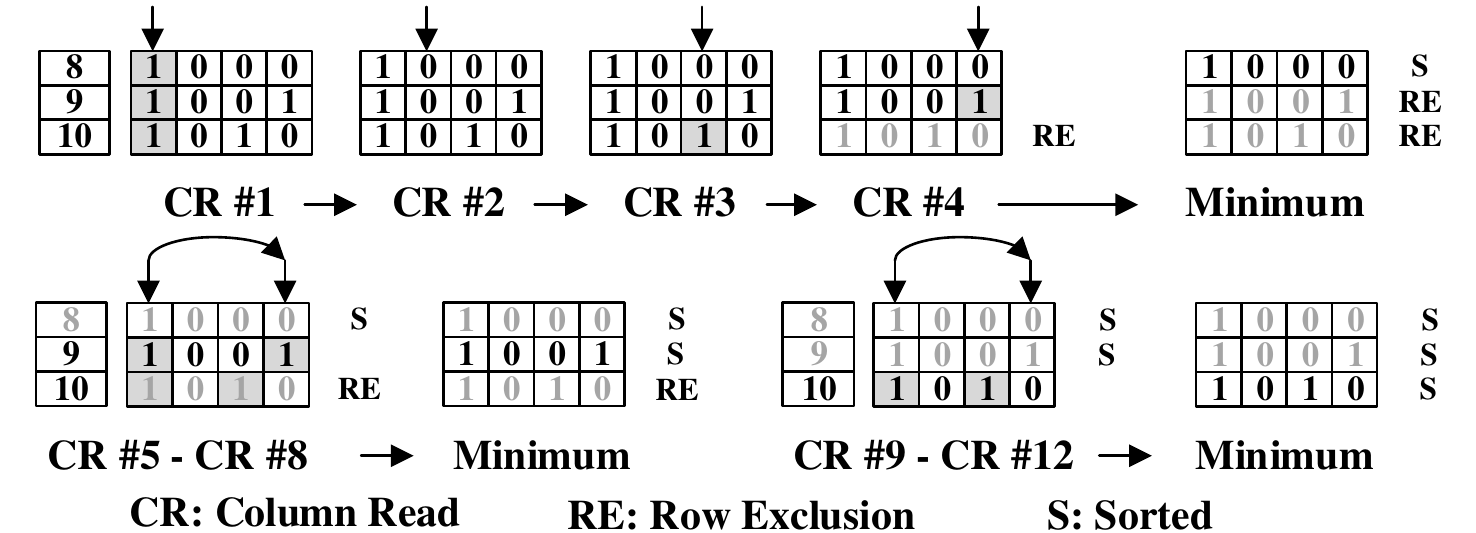}
    \caption{Memristive in-memory sorting \cite{prasad2021memristive}.}
    \label{fig:memristive_sorting}
\end{figure}

Iterative in-memory min computation is proposed in \cite{prasad2021memristive} for memristive in-memory sorting. It uses $N$ iterations to successively search and exclude the min values in a length-$N$ array. Suppose each memristor cell stores a bit in a 1T1R memristive memory. \figurename~\ref{fig:memristive_sorting} shows an example for a length-$N$ ($N = 3$) array of $w$-bit ($w = 4$) numbers, \{8, 9, 10\}. 

In each iteration, a $w$-step bit traversal algorithm searches the min value: at step $j$ ($j = w-1 \rightarrow 0$), a near-memory circuit reads an bit column corresponding to the $j$-th bits of all array elements, searches for 1's in that bit column, and exclude the rows that have 1's. When a bit column contains all 0's or 1's, the row exclusion can be skipped. Rows that are corresponding to non-minimum values are excluded step by step until the min value is reached. The row for the min value is then excluded and marked as sorted before moving to the next min search iteration. The near-memory circuit is designed to support two operations, column read (CR) and row exclusion (RE), and their associated control logic. \figurename~\ref{fig:memristive_sorting} shows the steps to sort \{8,9,10\} using memristive in-memory sorting \cite{prasad2021memristive}. Note that the near-memory circuit in \cite{prasad2021memristive} does not keep track the number of remaining elements in the array; therefore it takes $N = 3$ iterations of min search, each contains $w = 4$ CRs. The total sorting latency is $N\times w = 12$ CRs. 

\section{Column-Skipping Memristive In-Memory Sorting}

We observe that memristive in-memory sorting in \cite{prasad2021memristive} introduces a large number of redundant CRs which are repeatedly executed on leading 0's or bit columns that have been processed previously. As shown in \figurename~\ref{fig:memristive_sorting}, when searching the 2nd minimum number 9, the first 3 CRs have been processed in the 1st iteration and are repeated in the 2nd iteration. To efficiently skip these redundant CRs, we propose a low-latency column-skipping algorithm. We use unsigned fixed-point number as example, but it can easily be applicable to signed fixed-point and floating-point number formats with small changes as described in \cite{prasad2021memristive}.

\subsection{Low-Latency Column-Skipping Algorithm}

Redundant CRs can happen in two scenarios: 1) array elements may include leading 0's. CRs on these leading 0's can be skipped at the beginning of each iteration; 2) some CRs may have been processed previously for REs, i.e., we do not need to exclude any new rows for those bit columns. 

\begin{figure}
    \centering
    \includegraphics[width = 0.80\linewidth]{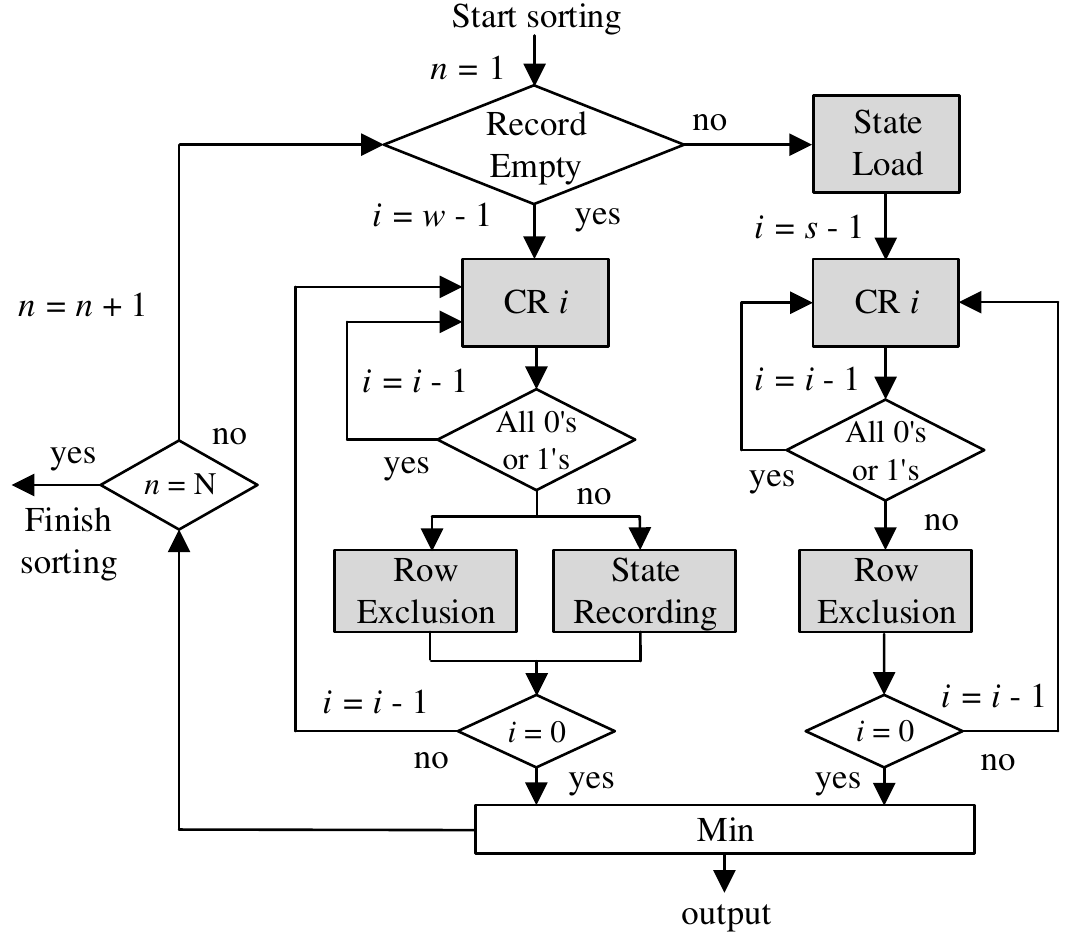}
    \caption{Iterative min search with proposed column-skipping algorithm}
    \label{fig:flow_chart}
\end{figure}

To detect and skip the redundant CRs, we propose to record the $k$ most recent RE states and their corresponding column indexes. The recorded states can be reloaded to skip redundant CRs. \figurename~\ref{fig:flow_chart} summarizes the iterative min computation for a length-$N$ array with proposed column skipping algorithm (where $n = 1 \rightarrow N$): 1) if state records are empty, the $w$-step algorithm \cite{prasad2021memristive} traverses each bit column from MSB ($i = w-1$) to LSB ($i = 0$). $k$ most recent RE states whose bit columns are not all 0's or 1's and their corresponding column indexes are stored in a state controller; 2) if state records are non-empty, we reload the most recent RE state and the corresponding column index $s$ and start from the next bit column $s-1$. CRs are executed on subsequent bit columns until reaching the min value.

\begin{figure}
\xdef\xfigwd{\textwidth}
\centering
        \includegraphics[width = 0.95\linewidth]{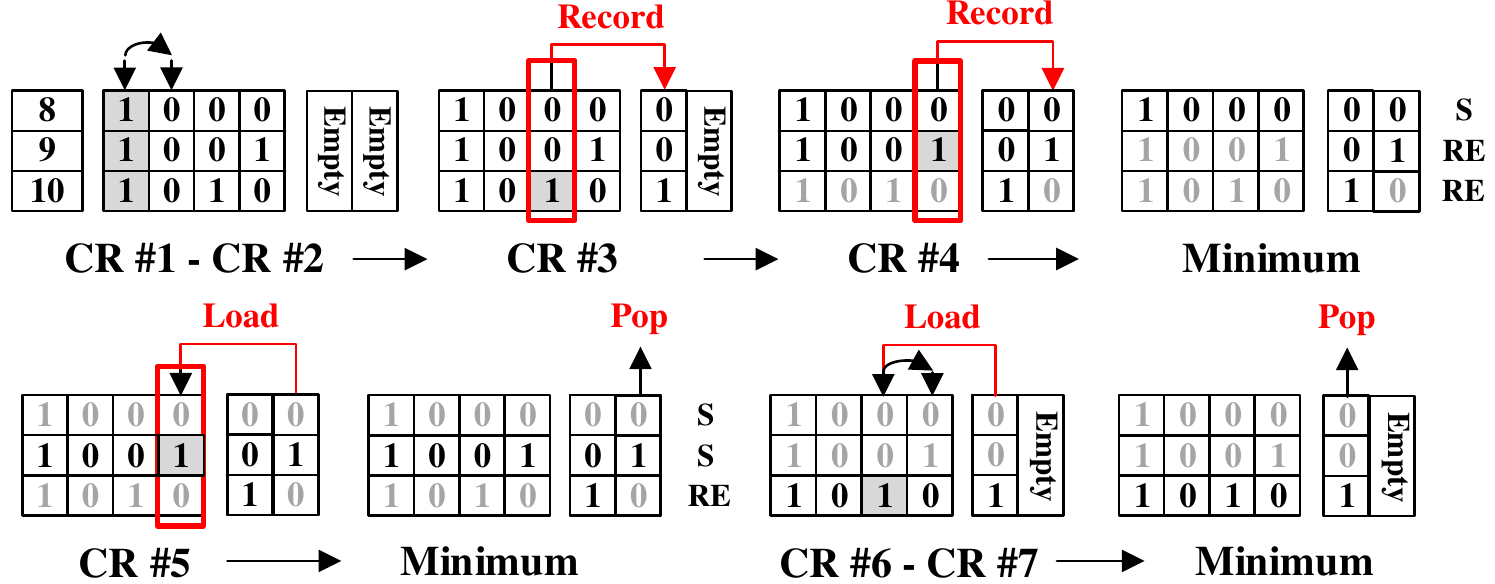}
    \caption{Column-skipping memristive in-memory sorting with state recording $k=2$.}
    \label{fig:memristive_sorting_k2}
\end{figure}

\figurename~\ref{fig:memristive_sorting_k2} illustrates the proposed column-skipping algorithm with state recording $k = 2$ when sorting the 4-bit array $\{8,9,10\}$. State recording in the first iteration helps to skip the first 3 CRs in searching the 2nd minimum and the first 2 CRs in searching the 3rd minimum. The total latency is reduced to only 7 CRs. The selection of $k$ affects the performance of the proposed column-skipping algorithm. We study the impacts of $k$ on sorting speedup, silicon area and power consumption in Section~\ref{section:evaluation}.  

\begin{figure}
    \centering
    \includegraphics[width = 0.91\linewidth]{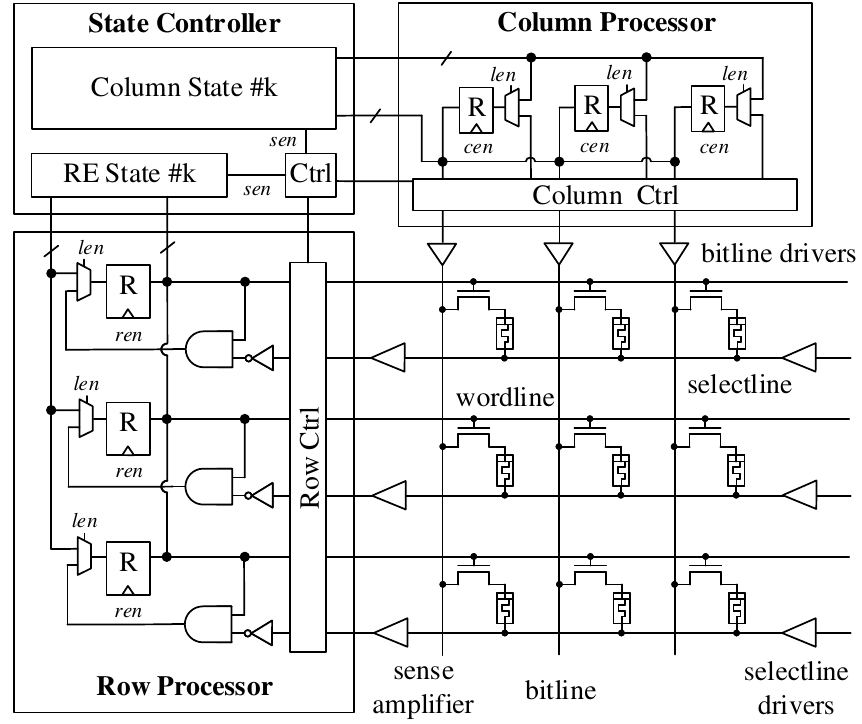}
    \caption{Near-memory circuit for column-skipping memristive in-memory sorting.}
    \label{fig:near_mem_circuit}
\end{figure}

\subsection{Near-Memory Circuit for Column-Skipping}

\figurename~\ref{fig:near_mem_circuit} demonstrates the near-memory circuit connected to a 1T1R memristive memory to implement the proposed column-skipping algorithm. The 1T1R memristive memory stores the binary bits of array elements with MSB on the leftmost column. Similar to \cite{prasad2021memristive}, select lines with sense amplifiers and bitline drivers are used for column reads. The proposed near-memory circuit consists of three modules: 1) a column processor that controls the column states; 2) a row processor that controls wordline (or RE) states; 3) a state controller that stores the RE states and their corresponding column indexes using a $k$-entry table. It also controls signals to execute all the operations.  

The near-memory circuit supports the four operations in \figurename~\ref{fig:flow_chart} as following: 1) column read (CR), where the column processor enables the bitline driver of a column and the corresponding bit column is read to the row processor. The column controller generates the next-step column state and the enable signal for column update ($cen$). Sense amplifiers measure the current on each select line to determine if it's 0 or 1; 2) row exclusion (RE), where the row processor checks if the bit column are all 0's or 1's (through row controller) before updating the wordlines (or RE) states. The row controller generates the enable signal for wordline update ($ren$). The wordlines that are connected to 1's are excluded and set to 0; 3) state recording (SR), where RE states and their corresponding column indexes are stored in a $k$-entry table. The recording is enabled ($sen$) if an iteration starts from the MSB and the bit column is not all 0's or 1's; and 4) state loading (SL), where the most recent RE state and the corresponding column index are sent to the row processor and column processor, respectively. The load enable signal ($len$) selects the reloaded states when updating the wordline and column registers. A top-level controller is used to schedule the four operations. 

When multiple rows remain unexcluded at the end of an iteration due to repetitions in the array, the column processor stalls to avoid redundant CRs until all repetition elements are excluded successively in the row processor.

\section{Multi-Bank Management}
\label{section:scalability}

\begin{figure}
    \centering
    \includegraphics[width = 0.95\linewidth]{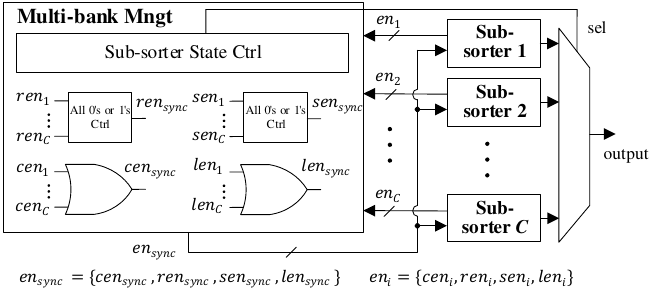}
    \caption{Multi-bank management to synchronize sub-sorter operations and select sorted output}
    \label{fig:scalability}
\end{figure}

The near-memory circuit shown in \figurename~\ref{fig:near_mem_circuit} can be scaled up to support larger array (i.e. larger $N$) or higher precision (i.e. large $w$). However, practical array can be too big to fit in a single memristive memory. To solve this problem, we propose a scalable solution to sort larger array stored in multi-bank memristive memory. 

Suppose a length-$N$ array is stored in $C$-bank memristive memory, each bank stores $N/C$ elements and has its own near-memory circuit that forms a length-$N/C$ sub-sorter. To realize length-$N$ sorting using $C$ sub-sorters of length-$N/C$, sub-sorters' operations need to be synchronized and run as a whole. A multi-bank manager is designed to connect the sub-sorters for this synchronization purpose: the judgement about all 0's or 1's needs to be considered globally to synchronize RE and SR operations while CR and SL operations are synchronized through the OR gates. 

\figurename~\ref{fig:scalability} shows the multi-bank manager to generate synchronized operation bits $en_{sync}$ based on local operation bits $en_i$ from sub-sorter $i$, where $i \in [1,C]$. In each sub-sorter, the synchronized operation bits $en_{sync}$ are used for replacing the original signals ($en_i$) to realize the corresponding function. The multi-bank manager monitors the sub-sorters' states and select the output from one of the $C$ sub-sorters if existing repetitions. Performance of the proposed multi-bank management are evaluated in Section~\ref{section:evaluation}. 

\section{Evaluation and Benchmarking}
\label{section:evaluation}

We evaluate the proposed techniques using statistically distributed datasets (uniform, normal and clustered) and practical datasets (from Kruskal's and MapReduce). We use 32-bit precision: the uniform distribution ranges from 0 to $2^{32}-1$, the normal distribution has a mean of $2^{31}$ and a standard deviation of $2^{31}/3$, and the clustered distribution has 2 clusters centered at $2^{15}$ and $2^{25}$ with identical standard deviation of $2^{13}$. To estimate the silicon area and power consumption, prototype sorters of length-1024 are implemented with 1T1R memristive memory using a 40nm CMOS technology. The RRAM device has two states and the corresponding resistances are 10M$\Omega$ and 100k$\Omega$, respectively. State-of-the-art memristive in-memory sorter \cite{prasad2021memristive} (baseline) and conventional digital merge sorter are implemented for comparison. All prototype sorters run at a 500MHz clock frequency.

\subsection{Sorting Speedup}

\begin{figure}
    \centering
    \includegraphics[width = 0.95\linewidth]{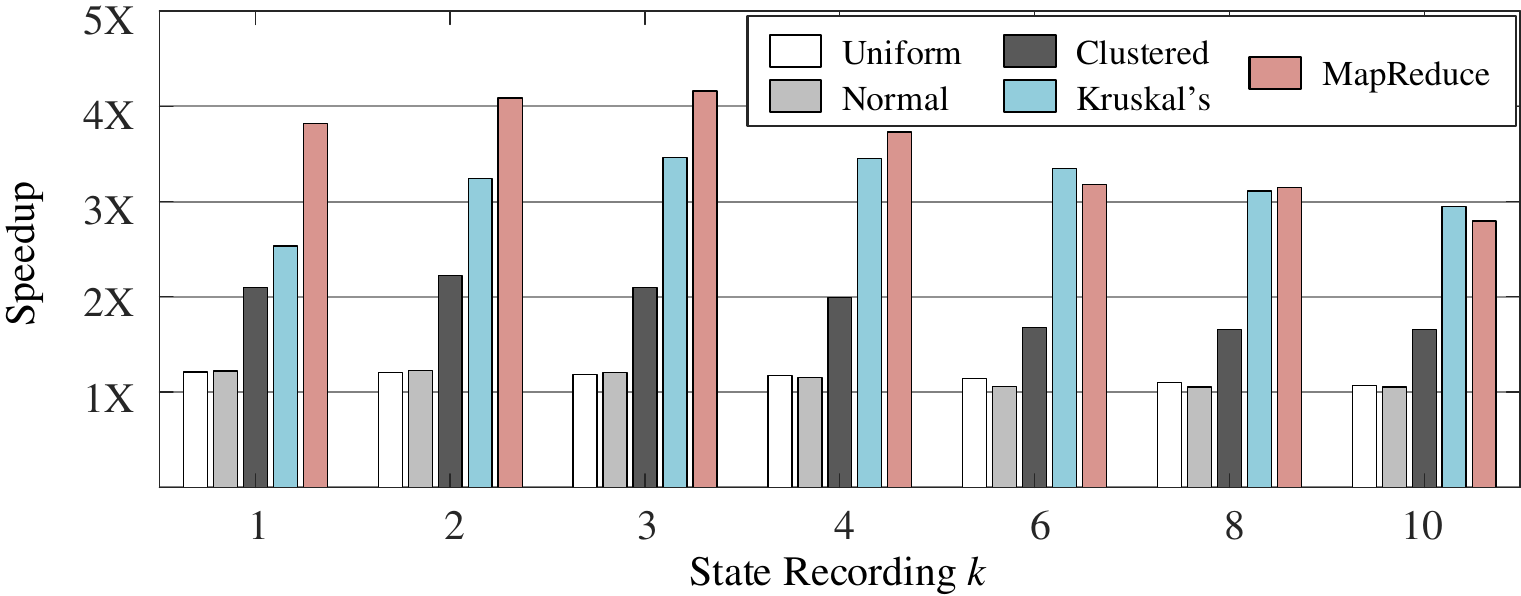}
    \caption{Normalized speedup over the baseline on different datasets with $N$ = 1024, $w$ = 32 and varying state recording $k$.}
    \label{fig:speed}
\end{figure}

The baseline implementation \cite{prasad2021memristive} has a fixed sorting speed of 32 cycles per number for any datasets. The merge sorter outperforms the baseline by 3.2$\times$ in speed. The speed of column-skipping sorter depends on parameter $k$ and dataset distribution. \figurename~\ref{fig:speed} shows the normalized speedup over the baseline on the selected datasets with $N = 1024$, $w = 32$ and varying state recording $k$. When $k$ increases, the min search is more likely to start from a recorded RE state; however, the reloaded starting position ($s$ in \figurename~\ref{fig:flow_chart}) may be further away from the optimal starting position, degrading the speedup due to less number of skipped CRs. We observe that the speedup saturates when $k$ reaches 2 or 3 and then goes down across selected datasets. 

The proposed column-skipping algorithm achieves faster sorting speed (up to 2.22$\times$ over the baseline) on clustered dataset than the speedup on uniformly or normally distributed datasets (up to 1.21$\times$ and 1.23$\times$ over the baseline, respectively). This is because clustered elements with small centers signify more leading 0's and redundant CRs. In Kruskal's and MapReduce dataset, majority of the small and repetitive elements lead to much better results for a speedup up to 3.46$\times$ and 4.16$\times$ over the baseline, respectively.  

\begin{figure}
    \centering
    \includegraphics[width = 0.95\linewidth]{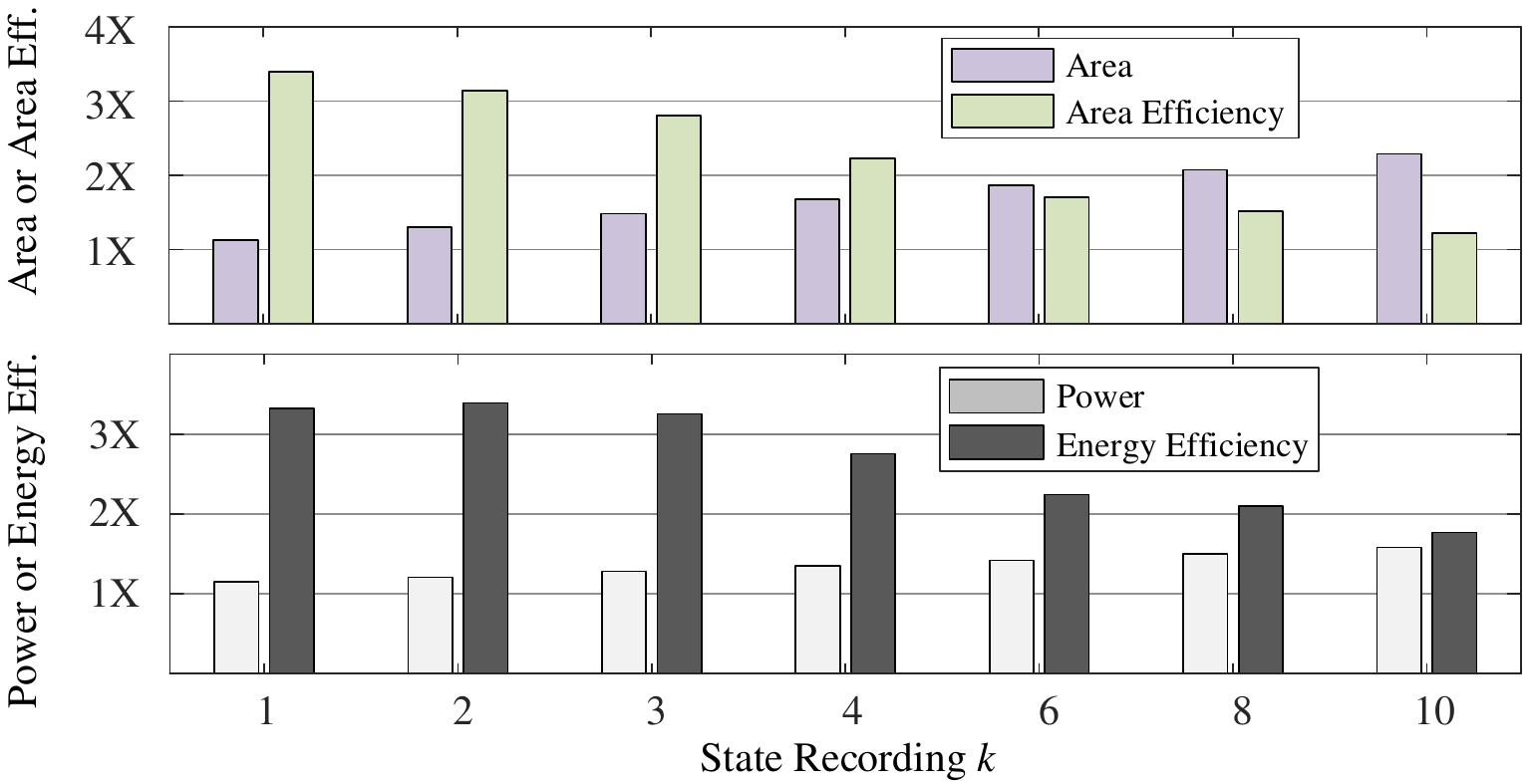}
    \caption{Normalized area and power over the baseline on MapReduce dataset with $N$ = 1024, $w$ = 32 and varying state recording $k$.}
    \label{fig:area_power}
\end{figure}

\subsection{Area and Energy Efficiency}

With $N = 1024$ and $w = 32$, the baseline sorter occupies 77.8K \textmu m$^2$ in silicon while the merge sorter occupies 246.1K \textmu m$^2$. The merge sorter demonstrates 1.01$\times$ area efficiency (throughput/area) over the baseline. We further measure the areas of column-skipping sorters with varying state recording $k$. \figurename~\ref{fig:area_power} presents the normalized area and area efficiency over the baseline when sorting the MapReduce dataset. With $k = 1$, column-skipping sorter demonstrates more than 3.2$\times$ area efficiency over the baseline. When $k$ increases, the sorter area increases due to larger state controller to store more RE states; however, the area efficiency goes down, because the speedup starts saturating when $k$ reaches 2 or 3. 

\begin{figure}
    \centering
    \includegraphics[width = \linewidth]{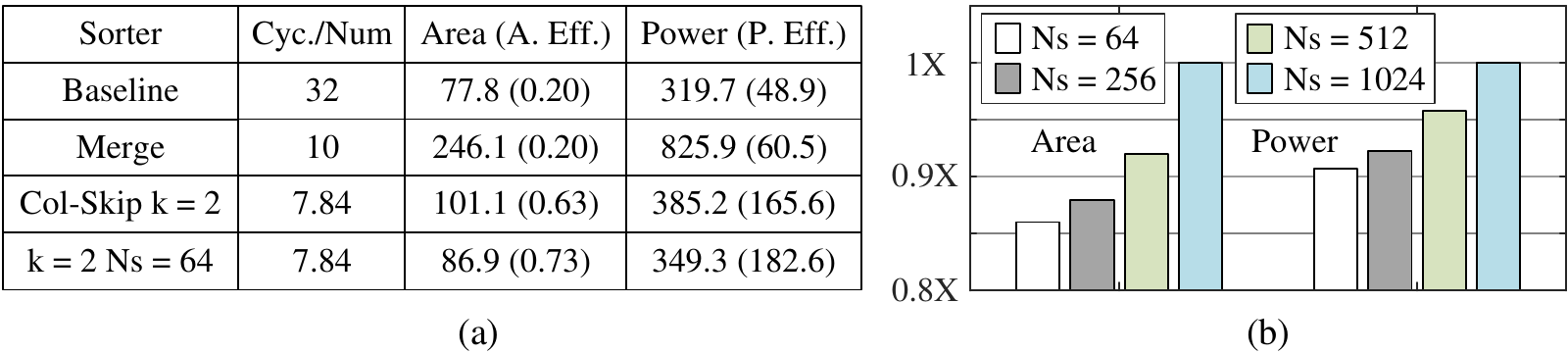}
    \caption{(a) Implementation summary using 40nm CMOS technology and 1T1R memristive memory (K \textmu m$^2$ for area, Num/ns/mm$^2$ for area efficiency, mW for power, Num/\textmu J for energy efficiency);(b) Normalized area and power (for MapReduce dataset) with varying sub-sorter length $N_s$ for $N$ = 1024, $w$ = 32 and $k = 2$ ($N_s = 1024$ is the baseline).}
    \label{fig:performance}
\end{figure}

We measured power using Ansys PowerArtist considering switching activities when sorting MapReduce dataset. The baseline sorter and the merge sorter consume 319.7 mW and 825.9 mW, respectively. The merge sorter demonstrates 1.24$\times$ energy efficiency (throughput/power) over the baseline. Column-skipping sorter consumes more power with increasing $k$, but the energy efficiency reaches the peak at $k = 2$ as shown in \figurename~\ref{fig:area_power}, outperforming the baseline by 3.39$\times$. The area and power consumption of 1T1R array are orders of magnitude less than the near-memory circuit. One can select the parameter $k$ based on target dataset for optimized speed, area and energy efficiency. 

\subsection{Multi-Bank Management}

To evaluate multi-bank management, we build a column-skipping sorter of $N = 1024$ using sub-sorters of length $N_s$ = 64, 256, 512. Multi-bank management does not change the speedup brought by column-skipping when clock frequency remains unchanged. Further reducing the sub-sorter length results in a degraded clock frequency under 500MHz due to more complex multi-bank manager. \figurename~\ref{fig:performance}(a) demonstrates the normalized area and power of multi-bank management over the original $N=1024$ sorter. We observe that the area and power of the near-memory circuit in sub-sorters decreases super-linearly when $N_s$ decreases. Even with an extra multi-bank manager, the total area and power for multi-bank management goes down with smaller sub-sorter length. Using 16 sub-sorters of length $N_s = 64$, the area and power reduction can be up to 14\% and 9\% compared to the original $N = 1024$ sorter. \figurename~\ref{fig:performance}(b) summarizes the implementation results for different sorters.

\section{Conclusions}

We present a fast and scalable memristive in-memory sorting that employs a column-skipping algorithm and a multi-bank management. Near-memory circuit with state recording is designed to efficiently skip redundant column reads for improved sorting speed and hardware efficiency. The multi-bank manager enables column-skipping for dataset stored in different banks of memristive memory. Prototype sorters are implemented using 40nm CMOS technology and 1T1R memristive memory. Experimented on a variety of sorting datasets with array length-1024, data precision 32-bit and state recording of 2, the speed, area efficiency and energy efficiency are 4.08$\times$, 3.14$\times$ and 3.39$\times$, respectively, than the state-of-the-art memristive in-memory sorting. 

\bibliographystyle{IEEEtran}
\bibliography{IEEEabrv,iscas2022}

\end{document}